# Synthetic Spatiotemporal Plasmonic Vortices On Chip

Qian Chen,[1,2,*] Shuoshuo Zhang,[2,3,*] Guoyu Xian,[4] Haoqiang Hu,[1] Xiaohua Wu,[1] Xiaofei Wu,[5,6] Jer-Shing Huang,[5,6] Chen-Bin Huang,[7] Jin-Hui Zhong,[8] Yuquan Zhang,[2] Xiaocong Yuan,[2] Changjun Min,[2,†] and Yanan Dai[1,9,‡]

[1]*Department of Physics, State Key Laboratory of Quantum Functional Materials, and Guangdong Basic Research Center of Excellence for Quantum Science, Southern University of Science and Technology, Shenzhen 518055, China*
[2]*Nanophotonics Research Center, Institute of Microscale Optoelectronics & State Key Laboratory of Radio Frequency Heterogeneous, Shenzhen University, Shenzhen 518060, China*
[3]*Key Laboratory of Light Field Manipulation and Information Acquisition, Ministry of Industry and Information Technology, School of Physical Science and Technology, Northwestern Polytechnical University, Xi'an 710129, China*
[4]*Beijing National Laboratory for Condensed Matter Physics, and Institute of Physics, Chinese Academy of Sciences, Beijing 100190, P. R. China*
[5]*Research Department of Nanooptics, Leibniz Institute of Photonic Technology, Albert-Einstein Str. 9, D-07745 Jena, Thuringia, Germany*
[6]*Institute of Physical Chemistry and Abbe Center of Photonics, Friedrich-Schiller-University Jena, D-07737 Jena, Thuringia, Germany*
[7]*Institute of Photonics Technologies, National Tsing Hua University, Hsinchu, Taiwan*
[8]*Department of Materials Science and Engineering, Southern University of Science and Technology, Shenzhen 518055, China*
[9]*Quantum Science Center of Guangdong–Hong Kong–Macao Greater Bay Area (Guangdong), Shenzhen 518045, China*
(Dated: November 11, 2025)

Spatiotemporal vortices are polychromatic modes that intertwine orbital angular momentum (OAM) in space and time. Here we introduce a new class of such vortices, *spatiotemporal plasmonic vortices* (STPVs), carrying nontrivial topological spin textures. They are generated by chronotopic interference of temporally delayed plasmonic eigen-vortices, where a $\pi$-phase dislocation in the space–frequency domain maps into a $2\pi$ spiraling phase in space–time, with the resulting focus–defocus dynamics emulate $U(1)$ gauge transitions. Using interferometric time-resolved photoemission electron microscopy (ITR-PEEM), we directly image their nanometer–attosecond (nano-atto) evolution and control vortex number and position. Quantum-path analysis of coherent two-photon photoemission (2PP) processes reveals the nonlinear plasmonic polarization fields and angular-momentum conservation, establishing STPVs as a platform for probing spatiotemporally structured quantum matter.

*Introduction*—Phase singularities, defined as points where the amplitude of a wavefield vanishes and the phase becomes undefined [1, 2], are universal topological defects independent of the microscopic details of the system. They are characterized by quantized angular momentum (AM), evolving robustly against perturbations and appearing across physics as vortices in superfluids and superconductors [3, 4], dislocations and disclinations in crystals [5], and skyrmions in magnetic textures [6]. In optics, they underpin OAM as a complement to spin angular momentum (SAM), which together define the vectorial degrees of freedom of light [7, 8]. In the paraxial regime, these quantities remain collinear with the propagation vector, but more generally they form complex textures due to spin–orbit interactions (SOIs) of light [9]. This gives rise to photonic merons, skyrmions, and related structures [10–17], linking optical vortices to defect classifications in condensed matter and field theory, providing concrete routes for manipulating light-matter interactions at the subwavelength scale.

Embedding singularities into the joint space–time domain introduces free-space spatiotemporal optical vortices (STOVs) as a new category of topological defects, representing genuinely polychromatic, non-separable solutions of Maxwell's equations [18–20] distinct from monochromatic or purely spatial fields. They can emerge spontaneously in the collapse and self-arrest of intense femtosecond pulses in dispersive media, or be deterministically sculpted by imprinting a spiral phase in the space-time or space-frequency domain using two-dimensional pulse shapers or nanophotonic metasurfaces. Once generated, STOVs carry tunable vectorial OAM and unconventional energy-flow, enabling phase-sensitive strong-field interactions [21] and nonlinear frequency conversion where spatiotemporal OAM is conserved up to extreme ultraviolet high harmonics [22, 23].

Achieving STOVs beyond propagating waves in extended media and integrating them into nanophotonic platforms holds great promise for scalable photonics and quantum simulators on the nano–atto scale [24–26]. A natural roadmap is to employ interface-bound modes such as surface and cavity polaritons, which offer strong field enhancement, deep subwavelength confinement, and compatibility with chip-based architectures [27–29]. In reduced dimensions, the space–time entanglement of AM can be further modified by boundary conditions and symmetry restrictions, pointing to unconventional classifications of topological defects and novel forms of photonic SOIs. Yet, dimensional reduction from volumetric beams to quasi-two-dimensional guided modes requires a conceptual advancement from free-space spatiotemporal engineering strategies, making the realization of true on-chip STOVs highly nontrivial.

---

[*] These authors contributed equally to this work
[†] Contact author: cjmin@szu.edu.cn
[‡] Contact author: daiyn@sustech.edu.cn

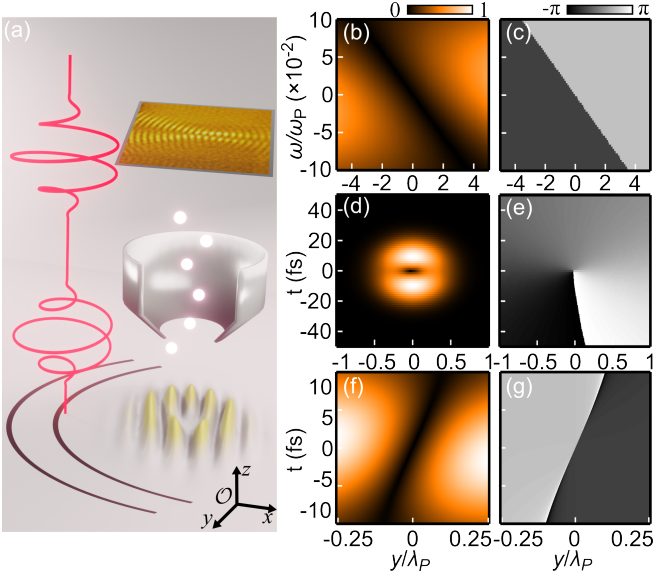

FIG. 1. Design principles of STPVs. (a) Schematic of STPV generation and ITR-PEEM imaging. Circularly polarized pulses are incident on a dual-spiral PCS with distinct geometric charges and an azimuthal separation (Figure S1). The resulting STPVs propagate and are focused by the PCS, while their interaction with the probe light induces 2PP of electrons for PEEM imaging. (b,c) Analytical model of space–frequency maps of amplitude and phase at the PCS plane, showing spatial chirp and a diagonal $\pi$-phase jump that signify space–frequency coupling, with $\Delta L_P = 2$ and $\Delta t_P = 1.5\lambda_P/c$, where $c$ is the local speed of light. (d,e) Fourier-transformed space–time maps, revealing a donut-shaped STPV with spiraling phase. (f,g) Focused HG-like field and phase distribution at the focal point $O$.

Here, we introduce a general synthesis approach, i.e. chronotopic engineering, which unifies geometric wavefront shaping with ultrafast spectral control [30, 31], to realize a family of STPVs on metallic surfaces. Distinct from conventional monochromatic vortex eigenmodes with spatially pinned cores [12, 13, 32–35], our approach launches a pair of surface plasmon polariton (SPP) wave packets with distinct OAM and a controlled temporal delay, whose interference enforces a $\pi$-phase dislocation in the space–frequency domain that evolves during propagation into quantized circulation in space–time, emulating a $U(1)$ holonomy manifested as a Berry phase [1, 2]. Through coherent nonlinear photoelectron imaging in ITR-PEEM [36], we directly capture this evolution through the nonlinear polarization underlying the coherent 2PP process, rendering the gauge transport observable with nano-atto resolution, while simultaneously resolving angular-momentum transfer at the quantum-path level. These measurements establish STPVs as a physical platform for tracing gauge-like evolution in real space and time, and for exploring how spatiotemporal topology couples to electronic states at surfaces.

*Theoretical formulation of STPVs*—The principle of chronotopic engineering of STPVs and their ITR-PEEM imaging are schematically shown in Fig. 1(a). Circularly polarized femtosecond pulses excite a dual-spiral plasmonic coupling structure (PCS, Supplementary Materials), launching a pair of forward-propagating SPP wave packets carrying distinct OAM [32, 33]. When these wave packets interfere, their differential OAM $\Delta L_P$ and frequency-dependent inter-plasmon delay $\Delta t_P$ give rise to space–frequency coupling, manifesting as a spatial chirp with a characteristic $\pi$-phase dislocation in the space–frequency domain [Figs. 1(b), (c)]. Such chronotopic coupling is conveniently described by the location of the frequency-dependent destructive interference:

$$y_{\text{null}}(\omega) \approx \frac{\pi - \omega \Delta t_P}{\Delta L_P}, \quad (1)$$

where the spatial and temporal coordinates are coupled by the slope $\Delta t_P/\Delta L_P$, which directly determines the line-defect (Figure S2 and S3). In the space-time domain, the surface normal electric field of STPVs can be formulated via integrating the desired frequency components of monochromatic eigenmodes and Fourier transforming into the time domain, yielding:

$$E_P(y,t) \propto \left[\frac{t}{\sigma_t^2} + i\frac{\Delta L_P}{\Delta t_P}y\right] \exp\left[-\frac{t^2}{2\sigma_t^2} - \frac{y^2}{2\sigma_y^2} - i\omega t\right], \quad (2)$$

where $\sigma_y$ and $\sigma_t$ represent the spatial and temporal full width half maximum (FWHM) of the vortex pulse. The temporal coordinate $t$ is defined in the co-propagating frame relative to the vortex core (See Supplementary Materials), in direct analogy to free-space STOVs [19, 20]. Consequently, the linearly chirped fields wraps into a vortex field with a donut-shaped profile and spiraling phase in the space–time domain [Figs. 1(d),(e)], representing a holonomy that carries a line defect in space–frequency into a winding singularity in space–time [1, 2]. Clearly, the spatial chirp and the pulse width together determine the eccentricity of the STPV.

Moreover, the focusing action of the PCS then reshapes the field into Hermite–Gaussian (HG) like modes with vanishing OAM and a symmetry-enforced phase inversion at the focal point, accompanied by the re-emergence of a line defect [Figs. 1(f), (g)]. Beyond the focus, the STPV revives with reversed OAM due to parity inversion along the $y$-axis. This full propagation cycle provides a physical analogue of gauge transport, where dislocations map into vortices and are inverted through $U(1)$ transformation, establishing STPVs as a platform for simulating related gauge fields.

*ITR-PEEM imaging of the STPV dynamics*—To realize these concepts experimentally, we fabricated PCSs on 100 nm thick polycrystal Au films and illuminated them with phase-locked, circularly polarized, 20 fs near infrared pulses under normal incidence, generating SPP vortex wave packets of center wavelength $\lambda_P = 800$ nm. Nano-atto visualization of STPVs was achieved via imaging of photoelectrons produced by direct or delayed 2PP. In this process, photoelectrons can be emitted either directly by the incident light, by the plasmons, or by their quantum mixture, allowing access to both the fundamental and harmonic components of the STPV field through Fourier analysis (Supplementary Materials) [13, 35, 37].

By scanning the delay $\Delta t_L$ between phase-locked excitation pulses, we directly captured the nano–atto dynamics of an isolated STPV in case of $\Delta L_P = 2$ (SM video 1). A fractional PCS was used to highlight the forward propagation. The resulting 2PP signal near the focal point shows a classic interferometric two-photon correlation, with additional modulation around 80 fs arising from plasmon propagation [Fig. 2(a)]. Fourier analysis of the corresponding ITR-PEEM movie isolates the spatiotemporally varying fundamental harmonic field $\Omega_1$ of the STPV (SM video 2 and Figure S8), as exemplified by the respective wave packets launched from each spiral [Fig. 2(a)].

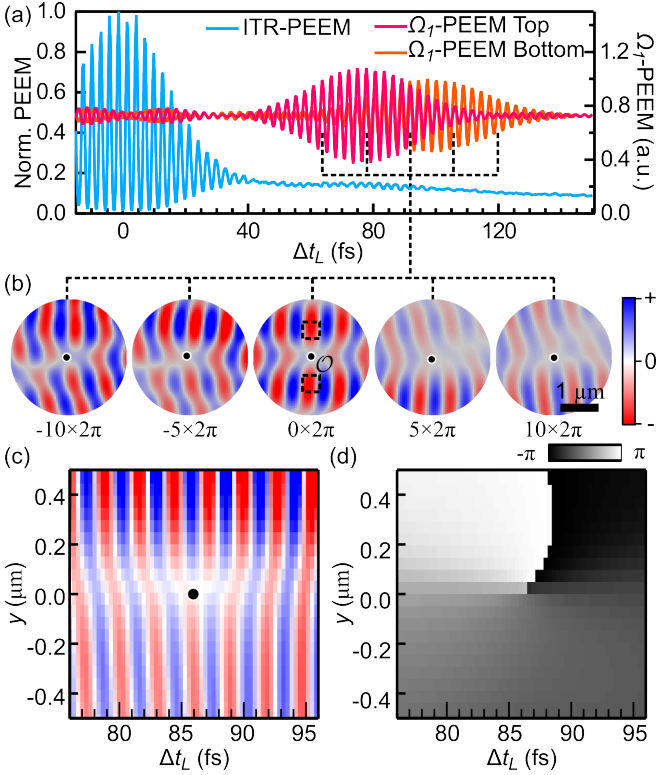

FIG. 2. Nano-atto imaging of STPVs. (a) Interferometric autocorrelation of normalized photoelectron yield within a $4\times4\lambda_P^2$ region around $\mathcal{O}$, together with the first-harmonic fields from the $1/2\times1/2\lambda_P^2$ subregions shown in (b). (b) Selected $\Omega_1$-PEEM snapshots of the propagating STPV, with time zero defined at the intersection of the two harmonics ($\sim$92 fs). (c,d) Amplitude and demodulated spiraling phase of the space–time reconstruction of the STPV. Black dots mark the vortex cores in all images.

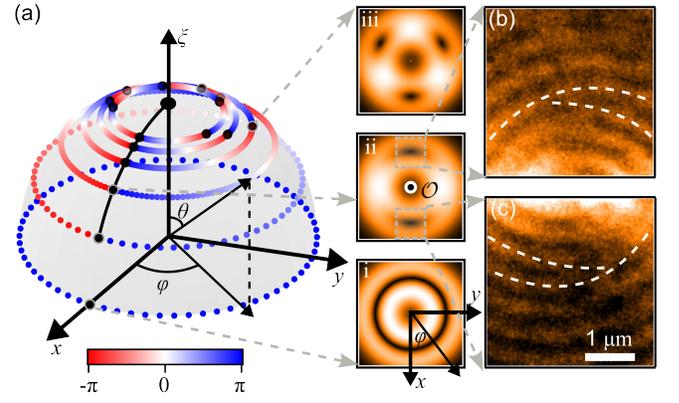

FIG. 3. Control of STPV generation. (a) Parametric map of STPV formation, showing vortex number determined by the $\Delta L_P$ and their azimuthal position tuned by the spiral separation $\Delta t_P$. Dotted circles correspond to integer values of $\Delta L_P$, with black dots indicating singularities when $\Delta t_P$ is a half-integer multiple of the plasmon period. Insets (i–iii) display simulated intensity distributions for 0,2,3 STPVs, illustrating how vortex number and azimuthal position can be tuned. (b),(c) Zoomed-in views of the upper and lower vortices, showing fork-shaped field distributions with two-fold rotational symmetry about $\mathcal{O}$.

Figure 2(b) presents a sequence of $\Omega_1$-PEEM images of the propagated STPV at different delay times. Before the focal point (negative delay), a characteristic fork-shaped distribution appears, demonstrating a spiraling phase dislocation (black dots) with topological charge of one. It corresponds to the SPTV right after being launched by the PCS [Fig. 1(d),(e)]. At point $\mathcal{O}$, the dislocation evolves back to a line-defect, where the STPV vanishes due to mirror symmetry and the spatial separation of the interfering plasmon packets. This corresponds to the HG-like mode shown in Fig. 1(f) and (g). Beyond this point, the vortex chirality reverses because of parity inversion along the $y$-axis, as shown by the reversal of the fork-shaped interference.

We further reconstruct the STPV evolution in the $y$–$t$ plane by slicing the time-dependent line profiles from $\Omega_1^-$PEEM imaging along the $y$-axis $\sim 4\lambda_P$ beyond the focal point, and stacking them sequentially in time, yielding the spatiotemporal field shown in Fig. 2(c). It unambiguously demonstrates a spatiotemporal vortex field distribution with the characteristic fork-shaped pattern in space-time. Moreover, with respect to the plane-wave-like fields farther away from the focal point $\mathcal{O}$, we phase demodulate the STPV in Fig. 2(c) to show its spiraling phase structure [Fig. 2(d)], confirming its gyrating field. We note that similar operations can be performed at different locations along the $x$-axis. While a spiraling phase can be observed for locations away from $\mathcal{O}$, HG-like mode is present near point $\mathcal{O}$, showing the continuous phase evolution between a line- and spiraling-phase defect (Fig. S9).

We emphasize that, unlike stationary plasmonic vortices whose cores remain fixed in space, STPVs originate from chronotopic coupling and are inherently broadband. More importantly, STPVs also support co-propagating meron-like topological spin textures similar to those in spatially pinned plasmonic vortices [11–14, 17], as shown by our finite difference time domain (FDTD) calculations in Fig. S10. They are created by joint action of the photonic SOIs and the vectorial fields of SPPs. Interestingly, however, the topologically nontrivial textures carried by STPVs evolve spatiotemporally and become trivial at the focal point. They further transit to nontrivial ones with opposite surface normal spin vector as the STPV defocuses and its OAM reverses. Such behavior naturally makes STPV as a nano-atto simulator of the topology evolution that is usually hidden in steady state measurements.

*Control over STPV generation*—The synthesis mechanism can be generalized to tune both the number and position of STPVs via $\Delta t_P$ and $\Delta L_P$, with the latter projected to $\xi = \cos\left[\frac{\pi}{2} - \tan^{-1}\left(\frac{\Delta L_P}{2}\right)\right]$ [Fig. 3(a)]. The $x$–$y$ coordinate corresponds to real space field distribution. In the space–time picture, no STPV forms when $|\Delta L_P| = 0$. This is similar to having a spatially pinned plasmonic vortex pulse featured by circular nodes [inset i of Fig. 3(a)] reported in the literature [12, 13, 32–35]. For $|\Delta L_P| = 1$, a unique phase singularity appears within the PCS, corresponding to a $2\pi$ dislocation and producing an isolated STPV with topological charge one. This is evidenced by the crescent shaped dark region in the Fig. S11(a), which is formed by collapsing the nodal ring of the Bessel function into a nodal point due to two-plasmon interference. As $|\Delta L_P|$ increases, higher-order dislocations fractionalize into multiple fundamental singularities equally spaced in azimuthal angle, and thus the number of STPVs grows accordingly. This effect arises because of the phase-slip and the rotational symmetry of the wave field, and the location of the singularity is given by:

$$\varphi_N = \frac{(2N+1)\pi - \omega_P \Delta t_P}{\Delta L_P}. \quad (3)$$

In addition, the azimuthal position of each STPV can be gradually tuned by adjusting $\Delta t_P$, as it determines the angle of nodal line and thus the propagation direction of STPVs. Such behavior is visualized by the blue–red gradient in Fig. 3(a), where a $2\pi$ phase accumulation occurs for each vortex, and the black dots mark the spatiotemporal phase singularities. Therefore, a $2\pi$ phase change in $\Delta t_P$ rotates the phase singularity by one angular period. As an example, Figs. 3(b)(c) show PEEM images of a pair of STPVs of charge one, formed with $\Delta L_P = 2$, but using a complete spiral PCS [Supplementary Materials]. This is different from the fractional PCS in Fig. 2, which only produces an isolated STPV, while leaving its mirror vortex not excited. In the present case, the two vortices appear symmetrically above and below the center $O$ [Fig. 3a(ii)]. Notably, they exhibit the same chirality with a clear two-fold rotational symmetry.

*Harmonics of STPV and OAM conservation*—Further insight of the generated spatiotemporal OAM lies in the analysis of the nonlinear harmonics of the ITR-PEEM movie. In case of 2PP processes, we can construct a three-level system $|i\rangle \rightarrow |m\rangle \rightarrow |f\rangle$, where electrons can reach the final state for photoemission through different quantum pathways involving time-ordered photon and plasmon interactions, as detailed in Supplementary Materials. Two examples of such pathways are shown in the double-sided Feynman diagrams of Figs. 4(a), (b) with photon and plasmon interactions indicated by the black and red arrows, respectively [35, 37, 38]. The left and right arrows represent the creation $\hat{a}^\dagger_{L,P}$ and annihilation $\hat{a}_{L,P}$ operators, respectively, and the vertical lines are time-ordered evolutions of the density matrix operators. Intermediate state population and coherence are not shown. The final state population $|f\rangle\langle f|$ rectifies into photocurrent that is imaged by PEEM. In static 2P-PEEM, all excitation routes contribute simultaneously, whereas in the interferometric mode, Fourier filtering isolates individual pathways (SM video S3-S5). The $\Omega_1$ component corresponds to the STPV field at the fundamental spatial-temporal frequencies arising from one-plasmon excitation [Fig. 4(a)], while the $\Omega_2$ compoenent originates from pathways with two successive plasmon interactions, producing the upconverted STPV field [Fig. 4(b)] at twice of the fundamental spatiotemporal frequency.

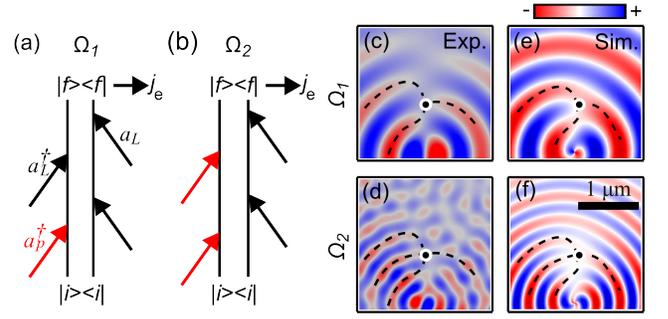

FIG. 4. STPV-induced nonlinear polarization fields. (a)–(d) Double-sided Feynman diagrams of the $\Omega_1$ and $\Omega_2$ components in a three-level system $|i\rangle \rightarrow |m\rangle \rightarrow |f\rangle$, illustrating the quantum pathways that generate the ITR-PEEM signal $j_e$. Black and red arrows denote photon ($\hat{a}^\dagger_L$, $\hat{a}_L$) and plasmon ($\hat{a}^\dagger_P$, $\hat{a}_P$) interactions, respectively. Detailed formulations are provided in the Supplemental Material. (c)-(d), Experimental PEEM images of $\Omega_1$ and $\Omega_2$ components, showing a first and second order OAM-driven field splitting, respectively. (e)-(f) Density matrix simulations of the STPV-induced polarization fields based on the quantum pathways in (a) and (b). Dashed lines and black dots in (c)-(f) mark the fork-shaped field distribution.

We unravel these pathways by Fourier analysis of the spatiotemporal dynamics of the STPV pair in Fig. 3. Zoomed-in snapshots of the $\Omega_1$- and $\Omega_2$-PEEM images when the vortices have propagated about $1\lambda_P$ from the focal point are shown in Fig. 4(c) and (d). The $\Omega_1$ image displays the characteristic fork-shaped distribution with topological charge one, while the $\Omega_2$ image reveals a clear second-order splitting, directly confirming the OAM up-conversion to charge two, and thus the OAM conservation. Figures 4(e), (f) present density-matrix simulations based on the quantum pathways of Figs. 4(a), (b), which closely reproduce the experimental results. Such second-order nonlinear polarization inherits the STPV topology and manifests as a $4\pi$ phase winding. Predictably, higher STPV harmonics also obey the OAM conservation, and thus embedding higher order topological defects.

*Discussions*—The STPVs discussed in this letter represent an exotic new class of surface excitations that can unravel gauge-field topology in real time. They are formed by the coherent superposition of temporally delayed plasmonic vortices with distinct OAM, such that a $\pi$-phase line-defect in the space–frequency domain Fourier-maps into a spiraling vortex defect in space–time. This evolution corresponds to a holonomy along a one-dimensional $U(1)$ curve, embedding a geometric Berry phase into the spatiotemporal domain [1, 2]. As the field propagates from the PCS to the focal point, the $2\pi$ vortex circulation unwraps into a discontinuous $\pi$-phase step and then reforms as a vortex beyond focus, directly enacting the same gauge holonomy that underlies the geometric phase, the Aharonov–Bohm effect, and the conversion of disclinations into quantized vortices. In addition to their vortex cores, STPVs support spin textures that resemble merons and skyrmions [11–14, 17], linking plasmonic fields to topological quasiparticles familiar from magnetism and condensed matter physics. In this sense, STPVs function as a robust chip-scale quantum simulator of topological phase evolution, transforming abstract gauge principles into directly observable electromagnetic field dynamics. More importantly, ITR-PEEM provides the crucial readout of this simulator, imaging not only nano-atto vortex dynamics but also the quantum pathways through which plasmons couple to photoemitted electrons, where competing processes can be disentangled by their spatial, temporal, and momentum signatures. While our present measurements focused on OAM conservation, the same approach can access intertwined OAM–SAM textures such as merons and skyrmions, and in the future may be extended to fractional angular momentum states and hybrid spin–orbit structures anticipated in dispersive media [39, 40].

These advances are enabled by chronotopic engineering, a geometric synthesis strategy that unifies spatial wavefront shaping with ultrafast spectral control directly at the plasmonic interface. In contrast to earlier demonstrations of spatiotemporal vortices that relied on bulky 4-$f$ systems [19, 20, 22], our approach is purely geometric, eliminating the need for pre-encoded pulse shaping or interferometric stabilization and requiring only free-space illumination. It is intrinsically robust and flexible: although circular polarization is used here for rotational symmetry, the position of the spatiotemporal singularity is set solely by interference and is therefore polarization-insensitive. This simplicity makes the method broadly transferable to other guided-wave platforms, including excitonic, phononic, and polaritonic modes. Beyond its procedural novelty, the intrinsic topology of STPVs also makes them exquisitely sensitive to weak perturbations such as dielectric inhomogeneity and delocalized Coulomb screening, positioning them as metrological probes for cavity fluctuations [41], high harmonic generation [22, 42, 43] and attosecond electron emission [44, 45]. Moreover, STPVs can imprint skyrmion or meron topological textures [25] onto a variety of quasiparticles, opening pathways toward spatiotemporally structured states of artificial quantum matter for on-chip information processing. In the strong-field regime, they further drive Floquet dynamics, connecting propagating dislocations to the quantum geometry of electronic wavefunctions [10, 46].

In conclusion, we have introduced chronotopic engineering as a general framework for generating spatiotemporal vortices in reduced-dimensional systems, and demonstrated their experimental realization through nonlinear photoelectron imaging of tunable, on-chip STPVs and their OAM-conserving harmonics with nano–atto resolution. Beyond establishing a new synthetic route for structured surface waves, our results highlight the topological nature of the underlying spatiotemporal defects and the power of quantum-path–sensitive imaging to access their dynamics. These findings provide a unifying paradigm for ultrafast nanophotonics and structured wave control, integrating topological wavefront design, strong-field light–matter interactions, and quantum information science within chip-based platforms.

*Acknowledgments*—This work was supported by the Ministry of Science and Technology of China (Grant Nos. 2024YFA1409800 and 2023YFA1407300), the National Natural Science Foundation of China (Grant Nos. 12374223 and 92477108), Guangdong Major Project of Basic and Applied Basic Research (No. 2020B0301030009), the Shenzhen Science and Technology Program (Grant Nos. 20231117151322001, KQTD20240729102026004, RCJC20210609103232046), Shenzhen University 2035 Initiative (Grant No. 2023B004), Research Team Cultivation Program of Shenzhen University (Grant No. 2023QNT014), and the Guangdong Provincial Quantum Science Strategic Initiative. Q.C. thank the SUSTech Public analysis and Testing Center for providing use of HIM instruments.